\documentclass[12pt]{article}

\begin{document}

\author{A.I.Volokitin$^{1,2}$ and B.N.J.Persson$^1$ \\
\\
$^1$Institut f\"ur Festk\"orperforschung, Forschungszentrum \\
J\"ulich, D-52425, Germany\\
$^2$Samara State Technical University, 443100 Samara,\\
Russia}
\title{Adsorbate induced enhancement of  
electrostatic non-contact friction }
\maketitle

\begin{abstract}
We study  the non-contact friction between an atomic force microscope tip and a metal 
substrate 
in the presence of bias voltage. 
The friction is due to energy losses in the sample
 created by the electromagnetic field from the oscillating 
 charges induced on the tip surface  by the bias voltage. 
We show that the friction can be enhanced by many orders of magnitude if the adsorbate 
 layer  can support acoustic vibrations.
The  theory predicts the magnitude and the distance dependence of friction 
 in a good agreement with  recent puzzling non-contact friction experiment \cite{Stipe}.We demonstrate that even an isolated adsorbate can produce high enough friction 
to be measured experimentally.  
\end{abstract}

A great deal of attention has been devoted to non-contact friction between
 an atomic force microscope tip and a substrate \cite
{Dorofeev,Gostmann,Stipe,Mamin,Hoffmann}. 
This is related to the
practical importance of non-contact friction for
ultrasensitive force detection experiments. The ability to
detect small forces is inextricably linked to friction via the
fluctuation-dissipation theorem. For example, the detection of single spins
by magnetic resonance force microscopy \cite{Rugar}, which has been proposed for
three-dimensional atomic imaging \cite{Sidles} and quantum computation \cite
{Berman}, will require force fluctuations to be reduced to unprecedented
levels. In addition, the search for quantum gravitation effects at short
length scale \cite{Arkani} and future measurements of the dynamical Casimir
forces \cite{Mohideen} may eventually be limited by non-contact friction
effects. Non-contact friction is also responsible for the friction drag force 
between two-dimensional quantum wells \cite{Gramila1,Gramila2,Sivan}. 

The non-contact friction
always is of electromagnetic  origin, but the detailed mechanism is not
totally clear, since there are several different mechanisms of energy
dissipation connected with the electromagnetic interaction between bodies.
First, the fluctuating current density in one body will induce a current 
density in other body. The interaction between the fluctuating current and 
the induced current density gives
rise to the well-known long-range attractive van der Waals interaction
 \cite{Lifshitz} and
 the van der Waals friction \cite{Volokitin1,
Volokitin2,Volokitin3}. Secondly, the surfaces of the tip and the sample always 
have static charges due to inhomogeneities of the surfaces. 
The presence of inhomogeneous 
tip-sample electric fields is difficult to
avoid, even under the best experimental conditions \cite{Stipe}. 
 The electric field
can be easily changed by applying a voltage between the tip and the
sample. The friction associated with 
moving  charges  will be denoted  as the \textit{ electrostatic friction}. 

The fluctuating electromagnetic field near the surface of the tip will 
produce a fluctuating force acting on the tip, resulting in Brownian 
motion of the cantilever. Thus, just as damping in liquids is intimately 
connected with random impacts and Brownian motion of small 
particles \cite{Einstein,Smoluchowski}, the non-contact friction is 
necessarily connected with the random forces that drive the cantilever 
Brownian motion.

Recently several groups have observed unexpectedly large long-range non-contact 
friction \cite{Dorofeev,Gostmann,Stipe,Mamin,Hoffmann}. 
The friction force $F$ acting on the tip was found to be  
proportional to the velocity $v$, $F=\Gamma v$.  
Stipe \textit{et.al.}\cite{Stipe}
observed non-contact friction effect between a gold surface and a gold-coated
cantilever as a function of tip-sample spacing $d$, temperature $T$, and
bias voltage $V$. For vibration of the tip parallel to the surface they
found $\Gamma (d)=\alpha (T)(V^2+V_0^2)/d^n$, where $n=1.3\pm 0.2,$ and $%
V_0\sim 0.2\,\mathrm{V.}$ At 295\textrm{K}, for the spacing $d=$ 100\textrm{%
\AA\ }they found $\Gamma =1.5\times 10^{-13}\mathrm{\,kgs}^{-1},$ 
 An applied voltage of 1
 V resulted in a friction increase of $\Gamma=3\times 10^{-12}$kg/s at 300 K 
and $d=20$nm.

In a recent Letter, Dorofeev \textit{et.al.} \cite{Dorofeev} claim that 
the non-contact friction effect observed in \cite{Dorofeev,Gostmann} is due to
Ohmic losses mediated by the fluctuating electromagnetic field . This result
is controversial, however, since the van der Waals friction for good conductors 
like copper has
been shown \cite{Volokitin1,Volokitin2,Volokitin3,Persson and Volokitin} to
be many orders of magnitude smaller than the friction  observed by Dorofeev \textit{%
et.al.}.  
Realizing that the van der Waals friction between good 
conductors 
is too small to explain the experimental data,   
 in \cite{Greffet} it was proposed that  the  van der 
Waals friction may be strongly enhanced  between the high resistivity mica substrate 
and silica tip. However 
 the mica substrate and silica tip were coated by gold films thick enough 
to  completely screen the electrodynamic interaction between underlying dielectrics. 
The large friction obtained in \cite{Greffet} is due to the extremely small 
conductivity of mica ( $\sigma \sim 10^{-8} (\Omega$ m)$^{-1}$).   

At small separation $d \sim 1$nm 
resonant
photon
tunneling between the  adsorbate vibrational modes on the surfaces of the tip 
and the sample  
can enhance the friction  by seven order of
magnitude in comparison with the good conductors surfaces  
  \cite{Volokitin2, Volokitin3}. 
However the distance dependence ($\sim 
1/d^{6}$) is stronger than 
observed experimentally.  

 Recently, a theory of 
noncontact friction was suggested where the friction arises from Ohmic losses 
associated with the  
electromagnetic field created by moving charges induced by the 
bias voltage \cite{Chumak}. 
In the case of a spherical tip this theory predict the same weak distance dependence 
of the friction as observed in the  experiment, but the magnitude of 
the friction is  many orders of magnitude smaller than the experimental value. 
At present there is no 
theory which can explain satisfactory both   
the magnitude and the distance dependence of friction observed in \cite{Stipe}.    

In this Letter we present a novel explanation of the puzzling experimental data 
reported in \cite{Stipe}. 
We suggest that the large long-range non-contact friction is due to the 
electromagnetic interaction  of the moving charges,   
 induced on the  surface of the tip by the bias voltage, with  acoustic vibrations in 
an adsorbate layer on the 
surface of the sample. In particular, for the Cs/Cu(001) system 
the experiment suggest the existence of an acoustical film-mode even for 
the very dilute phase ($\theta=0.08$) \cite{Senet}.
 The negligible value of the substrate corrugation for  Cs/Cu(001) 
explains why no preferential 
adsorption site has ever been observed for this system \cite{Meyersheim}.    

We begin by considering a model in which the tip of a metallic cantilever of length $L$ 
is a section of a cylindrical surface with radius of curvature $R$ (see Fig.1). 
The cantilever is 
 perpendicular to a flat sample surface, which  occupies the $xy$ plane, with the 
 $z$- axis 
pointing into the sample. The tip displacement $\mathbf{u}(t)=\hat{x}u_0e^{-i\omega t}$
is assumed to be  parallel to the surface (along the $x$ axis), which will be a 
good approximation when the oscillation amplitudes $u_0$ is sufficiently small. 
The cantilever 
width $b$, 
i.e., the size in the direction perpendicular to the $xz$ plane, is taken to be 
much larger than the thickness $c$ ($b\gg c$). 
In the case of 
a cylindrical tip geometry, the electric field induced by the bias voltage 
 is the same as that which would be produced in the vacuum region by two charged wires 
passing through points at $z=\pm d_1=\pm\sqrt{2dR+d^2}$, where $d$ is the separation 
between 
the 
cylinder and the sample surface \cite{Landau}. The wires have charges $\pm Q$ per unit 
length, $Q=CV$, where $V$ is the bias voltage, and   
\begin{equation}
C^{-1}=2\ln[(d+R+d_1)/R]
\end{equation}
 The vibration of the tip will produce an oscillating electromagnetic field which is the same 
as  that which would be produced in the vacuum region from an  oscillating charged wire located 
at $z=d_1$. The energy dissipation in the sample induced by the electromagnetic field 
of the oscillating wire is determined by integrating the Poynting's vector over 
  the sample surface. Taking into account that the energy dissipation per unit time
 must be equal to $2\omega^2\Gamma|u_0|^2$, we get the following 
formula for friction coefficient 
for motion of the cylindrical tip parallel to sample surface 
\begin{equation}
\Gamma_{\|}=\lim_{\omega\to 0}2Q^2b\int_0^{\infty}dqqe^{-2qd_1}
\frac{\mathrm{Im}R_p(\omega,q)}{\omega},  \label{one}
\end{equation}
where $R_p$ is the reflection coefficient for $p$-polarized electromagnetic waves.           

The reflection  coefficient $R_{p}$, which
take into account the contribution from an adsorbate layer, can be obtained using an 
approach which was proposed in  \cite{Langreth}. Using this approach we get:
\begin{equation}
R_{p}=\frac {1-s/q\epsilon+4\pi n_a[s\alpha_{\parallel}/\epsilon+q \alpha_{\perp}]
-qa(1-4\pi n_aq\alpha_{\parallel})}
{1+s/q\epsilon+4\pi n_a[s\alpha_{\parallel}/\epsilon-q \alpha_{\perp}]
+qa(1+4\pi n_aq\alpha_{\parallel})}, \label{two}
\end{equation}
where
\begin{equation}
s=\sqrt{q^2-\left(\frac{\omega}{c}\right)^2\epsilon},   \\
\end{equation}
and  $\alpha_{\parallel}$ and $\alpha_{\perp}$ are the
polarizabilities of adsorbates
in a direction parallel and normal to the surface,  respectively. $\epsilon=1+4\pi i
\sigma/\omega$
is the bulk  dielectric function, where $\sigma$ is a conductivity, and $n_a$ is
the concentration of adsorbates. 
 In comparison with the expression obtained in \cite{
Langreth}, Eq.(\ref{two}) takes into account that the centers of the adsorbates 
are located at distance $a$ away from image plane of the metal.  
Although  this gives corrections of the order $qa\ll 1$ to the reflection 
 amplitude, for parallel 
 adsorbate vibrations on  the good 
conductors (when $\epsilon\gg 1$), they  give the most important contribution 
to the energy dissipation. 
  For clean surfaces $n_a=0$, and in this
case formula (\ref{two}) reduces to the well-known Fresnel formula. In this case, 
 for $R\gg d$,  Eq.(\ref{one}) gives a formula which was obtained recently  in \cite{Chumak}
using a less  general approach:
\begin{equation}
\Gamma_{cl}^c = \frac{bV^2}{2^6\pi \sigma d^2}
\end{equation} 
With  $b=7\cdot10^{-6}$m and  $\sigma = 4\cdot10^{17}$s$^{-1}$ for gold at 300 K, 
and with $d=20nm$ and $V=1$Volt this formula gives a friction  
which is  eight orders of  magnitude smaller than observed experimentally \cite{Stipe}. 

Let us now consider ions with charge $e^*$ adsorbed on the sample.
The polarizability for ion vibration normal and parallel to the surface is given by
\begin{equation}
\alpha_{\perp(\|)}=\frac {e^{*2}}{M(\omega_{\perp(\|)}^2-\omega ^2 -
i\omega \eta_{\perp(\|)})},
\end{equation}
where $\omega_{\perp(\|)}$ is the frequency of the normal (parallel) 
adsorbate  vibration,
and $\eta_{\perp(\|)}$  the corresponding damping constant and $M$ is the adsorbate 
mass. A particular 
interesting case is 
Cs  adsorbed on Cu(100). At $\theta\approx 0.1$  the Cs  film has an acoustical 
branch with adsorbate vibrations parallel to the surface. This means that for such 
vibrations 
$\omega_{\|}\approx 0$. In this case the main contribution to the friction comes from 
vibrations parallel to the surface, and the imaginary part of the reflection 
amplitude is given by 
\begin{equation}
\mathrm{Im}R\approx\frac{2\omega \eta qa\omega_q^2} 
{(\omega^2-\omega_q^2)^2+\omega^2\eta^2}   \label{refl}
\end{equation}
where
\begin{equation}
\omega_q^2=\frac{4\pi n_ae^{*2}aq^2}{M} \label{three}
\end{equation}
Using (\ref{refl}) in (\ref{one}) for $R>>d$ we get
\begin{equation}
\Gamma_{ad}^c=\frac{b\eta MR^{1/2}V^2}{2^{7/2}d^{3/2}\pi n_ae^{*2}} \label{four}
\end{equation}
This friction exhibits the same distance dependence as is observed experimentally 
\cite{Stipe}. For  isolated Cs on Cu(100) the damping parameter $\eta$ 
has both phononic and electronic contributions.  
The friction due to  one-phonon processes   gives   
$\eta_{ph}=2.6\cdot 10^8$s$^{-1}$
 \cite{Senet}. However the existence of the acoustic branch in the 
adsorbed layer means that adsorbate layer is incommensurate with the substrate. 
In this case  emission of the bulk 
phonons gives vanishing contribution to the damping of parallel vibrations in 
the adsorbed layer because the adatoms do not``see" the corrugation of the 
substrate potential. 
 
For Cs adsorbed on Cu(100) the adsorption 
height $a=2.94${\AA} and at coverage $\theta\approx 0.1$ the dipole moment 
$\mu\approx4D$ \cite{Senet}, thus the ion charge $e^{*}=\mu/a\approx 0.28$e. In this 
case the theory  of the electronic friction for ionic bond  \cite{Persson,Liebsh}
gives 
$\eta_{\| ion}=3\cdot 10^7$s$^{-1}$. This is a rather small damping constant. 
Much larger 
contribution comes from electronic friction due to covalent bond which is given by 
\cite{Persson,Volokitin4}. 
\begin{equation}
\eta_{\| cov}=\frac{2}{\pi\hbar}\xi \frac{m_e}{M}\epsilon_F \sin^2\frac{\pi e^*}{e}
\label{cov}
\end{equation}
where for Cs adsorbed on Cu(100) the parameter $\xi\approx 0.17$, $m_e$ is the 
electron mass, $\epsilon_F$ is the Fermi energy. With $\epsilon_F=7$eV , $e^*/e=
0.28$, Eq.(\ref{cov}) gives $\eta_{\| cov}=3\cdot 10^{9}$s$^{-1}$ and 
with $n_a=10^{18}$m$^{-2}$ ($\theta \approx 0.1$), $R=10^3$nm, and with the same 
other parameters as above at $d=20$nm Eq.(\ref{four}) 
 gives $\Gamma=2.7\cdot 10^{-13}$ what is seven order of 
magnitude larger than for the clean surface, and only one order of magnitude smaller 
than observed experimentally 
\cite{Stipe}. Because of similarities of Cu and Au surfaces the same estimations 
will be valid also for the Au surface.  
 However the electronic friction predicts weaker temperature dependence \cite
{Persson1}
than in the experiment, where the friction at 300 K for $d=20$nm is approximately 
6 times larger than at 77 K regardless of voltage.  

Defects of the adsorbed layer like atomic steps, impurity atoms  
and other imperfections, and also the anharmonicity 
will also contribute to vibrational relaxation in adsorbed layer. 
The anharmonicity and vacancies give temperature dependent contribution to 
 the damping parameter $\eta$. The vacancies give exponential 
dependence on the temperature, and 
anharmonicity gives contributions $\sim T$ for three-phonon processes and 
$\sim T^2$ for four-phonon processes.  
Further experimental and theoretical studies will be required to fully elucidate 
the origin of the phonon life-time in adsorbed layer.

  In the case of
a spherical
 tip geometry  the electric field induced by the bias voltage 
 is approximately the same as that which  would be produced in the vacuum region 
between  two point charges $\pm Q =\pm CV$
located at 
\begin{equation}
z=\pm d_1=\pm\sqrt{3Rd/2+\sqrt{(3Rd/2)^2+Rd^3+d^4}}  \label{d1}
\end{equation}
 where 
\begin{equation}
C=\frac{d_1^2-d^2}{2d}  \label{C}
\end{equation}
It can been  shown that  the electrostatic force between the tip and the metal 
surface within this  approximation   agrees very well with  the exact expression 
 for a  sphere above metal surface \cite{Hudlet}. The 
vibrations of the tip 
will produce an oscillating electromagnetic field,  which in the vacuum region 
coincides 
with the electromagnetic field of an oscillating point charge.
 The friction coefficient for a point charge 
 moving parallel to the surface due to the electromagnetic energy losses 
inside the sample, 
is determined by  \cite{Persson and Schaich} 
\begin{equation}
\Gamma_{\|}=\lim_{\omega\to 0}\frac{Q^2}{2}\int_0^{\infty}dqq^2e^{-2qd_1}
\frac{\mathrm{Im}R_p(\omega,q)}{\omega}  \label{five}
\end{equation}
For motion normal to the surface, $\Gamma_{\perp}=2\Gamma_{\|}$. For a clean 
sample surface 
and for $R>>d$ from Eq.(\ref{five}) we get
\begin{equation}
\Gamma_{cl}^{s}=\frac{3^{1/2}R^{1/2}V^2}{2^7d^{3/2}\pi \sigma}
\end{equation}
This expression is only  a factor 1.6 smaller that  obtained independently 
in \cite{Chumak} using a less general approach. For  the same parameters as above 
and at $d=20$nm, the friction for a spherical tip is  two 
order of  magnitude smaller than for the cylindrical tip. For a spherical tip, 
 when the sample surface 
is covered by adsorbates,  from  Eqs.(\ref{refl}) and (\ref{five}) for $R\gg d$ 
we get 
the contribution to the friction from adsorbates  
\begin{equation}
\Gamma_{ad}^s=\frac{3RM\eta V^2}{2^6d\pi n_ae^{*2}}
\end{equation}
This  friction at $d=20$nm   is  also two order of magnitude 
smaller than for the cylindrical tip.

We now show that the atomic force microscope can be used  for 
to study the vibrational dynamic of isolated adsorbates. Let us calculate 
the friction acting on  the tip due to the interaction of the tip with vibrations of 
an isolated adsorbate. Let us assume that the tip is in a top position relative to an 
adsorbate. In this case, due to vibrations of the ion parallel to the surface,
 on the tip will act 
a fluctuating force in direction normal to the surface
\begin{equation}
F_z^f=\frac{12Q\mu u_{a\|}^2}{d_1^5} \label{force}
\end{equation}
where $u_{a\|}$ is the displacement coordinate for vibrations of the 
ion parallel to the 
surface, $\mu$ is the dipole moment of isolated ion, and $d_1$ and $Q$ 
are determined  by Eqs.(\ref{d1}) and (\ref{C}). Accordingly to 
the Kubo formula, the friction coefficient can be expressed through the force-force 
correlation function \cite{Schaich}
\begin{equation}
\Gamma_{\perp}=(k_{B}T)^{-1}\mathrm{Re}\int_0^\infty dt \langle F_z^f(t) F_z^f(0)\rangle
\label{Kubo}
\end{equation}      
where $\langle...\rangle$ stands for thermal equilibrium average. Using Eqs.
(\ref{force}) and (\ref{Kubo}) for $d\gg R$ we get 
\begin{equation}
\Gamma_{\perp}=1.3\frac{\mu^2 k_{B}TV^2}{M^2\omega_{\|}^4d^{5}R^3\eta}
\end{equation}
For the Cs/Cu(100) system $\mu=7.5$D, $\omega_{\|}=0.9\cdot 10^{12}$s$^{-1}$, 
$\eta=4.6\cdot
10^{9}$s$^{-1}$ \cite{Senet}. For a sharp tip with $R=0.5d=1$nm, $V=1$Volt 
and $T=300$K we get 
friction 
$\Gamma_{\|}=7.1\cdot 10^{-13}$kg/s. This friction is so large  that it can be measured
 with present state-of-the-art equipment.
  Note that the friction is characterized by a very strong distance dependence,
 and this can be used for the determination of the ion position.

In this letter, we have shown that the ``electrostatic friction'' can be enhanced
by several orders of magnitude when the surface of the metal sample is covered 
by an adsorbed layer with an acoustic branch of vibrations parallel to the surface. 
Our calculations predict a non-contact friction with similar magnitude, 
distance and bias voltage dependence as observed in a recent 
atomic force microscope study \cite{Stipe}. We have demonstrated 
that even isolated adsorbate 
can produce high enough friction which can be measured experimentally.
 These results should have broad  application in non-contact friction
microscopy, and  in the design of new tools for studying adsorbate vibrational dynamics.

\vskip 0.5cm \textbf{Acknowledgment }

A.I.V. acknowledges financial support  from the  Russian Foudation for 
Basic Research (Grant N 04-02-17606).
 B.N.J.P. and A.I.V. acknowledge 
support from the European Union Smart QuasiCrystals project. 
\vskip 1cm

FIGURE CAPTURE

Fig.1 Scheme of the tip-sample system. The tip-shape is characterized its 
length $L$ and the tip  radius of curvature $R$.

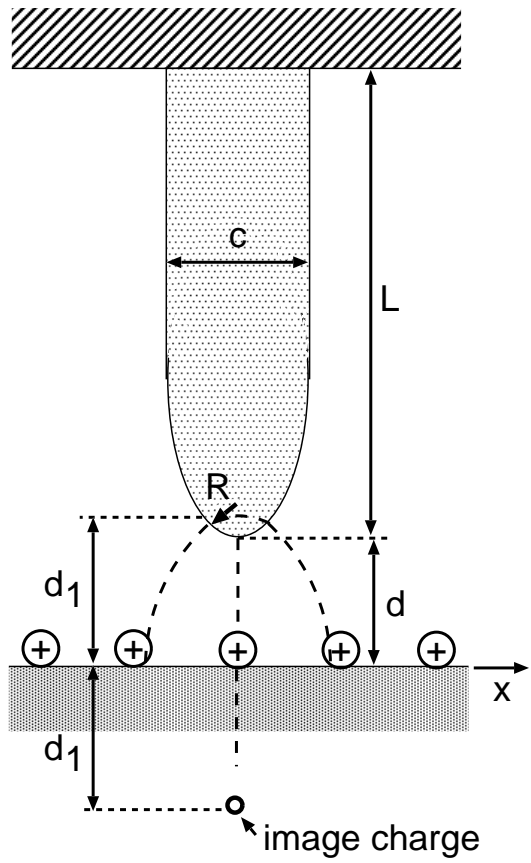
\end{document}